\newlength{\absize}
\documentclass[12pt]{article}
\usepackage{latexsym}
\usepackage{amssymb}
\usepackage{epsf}
\usepackage{setspace}
\renewcommand{\arraystretch}{2}

\newcommand{\figsize}{\small}

\input prepictex
\input pictex
\input postpictex
\newdimen\tdim
\tdim=\unitlength
\def\stpltsmbl{\setplotsymbol ({\small .})}

\newbox\sru
\setbox\sru=\hbox{\beginpicture
\setcoordinatesystem units <\tdim,\tdim>
\stpltsmbl
\setquadratic
\plot
  0.0   0.0
  4.8   1.5
  7.5   5.0
  7.3   8.5
  5.0  10.0
  2.7   8.5
  2.5   5.0
  5.2   1.5
 10.0   0.0
/
\endpicture}
\def\springru #1 #2 *#3 /{\multiput {\copy\sru}  at
#1 #2 *#3 10 0 /}
\newcommand{\np}{\mbox{\tiny$N$$+$$1$}}

\renewcommand{\bar}{\overline}
\usepackage{hyperref}

\newcommand{\cmt}[1]{{#1}}

\begin{document}

\thispagestyle{empty}
\pagestyle{empty}
\renewcommand{\thefootnote}{\fnsymbol{footnote}}
\newcommand{\starttext}{\newpage\normalsize
 \pagestyle{plain}
 \setlength{\baselineskip}{3ex}\par
 \setcounter{footnote}{0}
 \renewcommand{\thefootnote}{\arabic{footnote}}
 }
\newcommand{\preprint}[1]{\begin{flushright}
 \setlength{\baselineskip}{3ex}#1\end{flushright}}
\renewcommand{\title}[1]{\begin{center}\LARGE
 #1\end{center}\par}
\renewcommand{\author}[1]{\vspace{2ex}{\Large\begin{center}
 \setlength{\baselineskip}{3ex}#1\par\end{center}}}
\renewcommand{\thanks}[1]{\footnote{#1}}
\renewcommand{\abstract}[1]{\vspace{2ex}\normalsize\begin{center}
 \centerline{\bf Abstract}\par\vspace{2ex}\parbox{\absize}{#1
 \setlength{\baselineskip}{2.5ex}\par}
 \end{center}}

\preprint{\#HUTP-05/A0039}
\title{Chiral Fermion Delocalization in Deconstructed Higgsless Theories}
\author{
 Howard~Georgi,\thanks{\noindent \tt georgi@physics.harvard.edu}
 \\ \medskip
 Lyman Laboratory of Physics \\
 Harvard University \\
 Cambridge, MA 02138
 }
\date{July 2005}
\abstract{I construct a renormalizable $SU(2)^{89}\times U(1)$ 
gauge theory with
standard-model-like phenomenology for the gauge bosons masses and the weak
interactions of the light fermions (including the $b$) 
but in which all vacuum expectation
values are about $2$~TeV. This is a deconstructed version of a Higgsless
model with a flat extra dimension. 
The fermions are delocalized on the theory space in an unusual way, with
LH and RD fermions on alternate nodes.} 

\starttext


In this note, I report on an exercise in gauge theory model-building that
could have been done thirty years ago. The model is a conventional
renormalizable\footnote{But see the discussion in footnote~\ref{natural} on
page \pageref{natural}.} quantum
field theory with gauge symmetry spontaneously broken by the vacuum
expectation values of elementary scalar fields. The model is designed to
reproduce the electroweak interactions of the 
standard model to a good approximation in tree
approximation. The obvious
difference between this model and the conventional standard model with an
elementary Higgs boson is the size of the gauge group and the fermion
representation. The 
electroweak gauge group is $SU(2)^{89}\times U(1)$ and 
the number of fermion representations
is similarly swollen. 
The reason I consider this
ludicrously large structure is to illustrate 
in a very explicit way how a
so-called Higgsless model~\cite{Csaki:2003dt,Csaki:2003zu} 
can very nearly reproduce the low-energy
phenomenology of the standard model. 
In particular, the particle masses and the
couplings of the light fermions are within a percent or
two of their standard model values. Of course this model is not without
elementary Higgs bosons. However, it is a
deconstructed~\cite{Arkani-Hamed:2001ca,Hill:2000mu} version of a 
Higgsless model~\cite{Foadi:2003xa,Hirn:2004ze,Casalbuoni:2004id,
Chivukula:2004pk,Evans:2004rc,Georgi:2004iy,Chivukula:2004af,Kurachi:2004rj,
Chivukula:2004mu,Chivukula:2005bn}, and the translation of
``Higgslessness'' into the language 
of 70s model building is simply that the vacuum expectation values 
that break the $SU(2)$ symmetries are 
all much larger than the $250$~GeV of the standard
model. In the explicit example I will describe, the VEVs are all greater
than $2$~TeV. I find this rather remarkable and I simply could not have
imagined back in the 70s that such a thing was possible.

\cmt{It is quite clear in retrospect why models like this were not constructed
in the 70's. From the 4-dimensional point of view, these models look crazy,
with hundreds of apparently extra structures and parameters! What one gains
from thinking about extra dimensions is motivation to
consider models that look very complicated from the 4-d point of
view but which, from the 5-d point of view, are really rather simple. 
Indeed, in the explicit model I discuss here, I will 
make an assumption related to 
translation invariance in the extra dimension that reduces the number of
parameters to just three more than in the standard model. 
For me, this is all that extra dimensions are good for. I try to
deconstruct them as quickly as possible so that I know what I am doing. 
But I think perhaps 
that the youngsters who have been weaned on a geometrical picture gain
some intuition directly from 5-dimensional thinking.}

What is new in this note, I believe, is a more explicit and different
treatment of what is called in the 5-d language the ``delocalization'' of the
fermions. It is necessary to spread the the fermions out in the extra
dimensions to produce a negative contribution to the $S$
parameter~\cite{Cacciapaglia:2004rb,Foadi:2004ps,Casalbuoni:2005rs,
Chivukula:2005bn,Chivukula:2005xm}. 
In the
4-d language of this paper, this simply means that the light fermion
doublets are linear combinations of fields that transform under
different $SU(2)$ subalgebras. I will introduce a very efficient 
chiral fermion delocalization in which 
the left-handed and right-handed fermions appear on alternate nodes. I find
this a bit confusing from the 5-d point of view. More of this later.

I will begin by describing the model without much further motivation. Once
we have all the pieces and I have described a particular choice of
parameters that does what I want, I will step back and discuss things more
generally. 

The gauge structure of the model is summarized in pictorial form in
figure~\ref{fig-g1}. 
{\figsize\begin{figure}[htb]
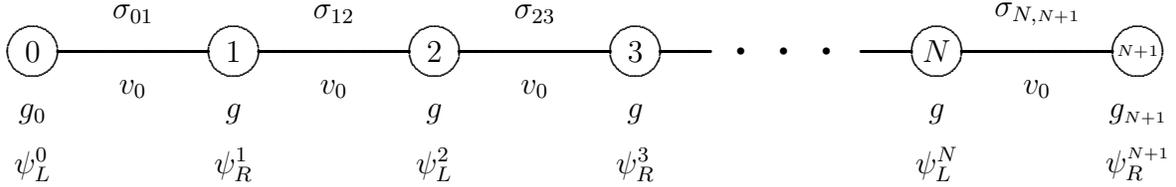

$$\beginpicture
\setcoordinatesystem units <.95\tdim,.95\tdim>
\circulararc 360 degrees from -170 0 center at -160 0
\circulararc 360 degrees from -90 0 center at -80 0
\circulararc 360 degrees from 10 0 center at 0 0
\circulararc 360 degrees from 90 0 center at 80 0
\circulararc 360 degrees from 210 0 center at 200 0
\circulararc 360 degrees from 290 0 center at 280 0
\put {$0$} at -160 0
\put {$1$} at -80 0
\put {$2$} at 0 0
\put {$3$} at 80 0
\put {\tiny$N$$+$$1$} at 280 0
\put {$N$} at 200 0
\stpltsmbl
\plot -90 0 -150 0 /
\plot -10 0 -70 0 /
\plot 10 0 70 0 /
\plot 110 0 90 0 /
\plot 170 0 190 0 /
\plot 210 0 270 0 /
\put {$\sigma_{01}$} at -120 15
\put {$\sigma_{12}$} at -40 15
\put {$\sigma_{23}$} at 40 15
\put {$\sigma_{N,\np}$} at 240 15
\put {$v_{0}$} at -120 -15
\put {$v_{0}$} at -40 -15
\put {$v_{0}$} at 40 -15
\put {$v_{0}$} at 240 -15
\put {$g_0$} at -160 -25
\put {$g$} at -80 -25
\put {$g$} at 0 -25
\put {$g$} at 80 -25
\put {$g_{\np}$} at 280 -25
\put {$g$} at 200 -25
\put {$\psi^0_L$} at -160 -45
\put {$\psi^1_R$} at -80 -45
\put {$\psi^2_L$} at 0 -45
\put {$\psi^3_R$} at 80 -45
\put {$\psi^{N}_L$} at 200 -45
\put {$\psi^{\np}_R$} at 280 -45
\multiput {\tiny$\bullet$} at 122 0 *2 17 0 /
\endpicture$$
\caption{\figsize\sf\label{fig-g1}The Moose diagram associated with the
example.}\end{figure}}  
There are $N+1$ $SU(2)$ factors in the gauge group, and $N$ is constrained
to be even, but is otherwise free. 
The links and nodes of the figure form a ``theory space''
that corresponds via deconstruction 
to the configuration space of the extra
dimension. 
Nodes $1$ through $N$ are associated with the ``bulk'' of the
extra dimension, between the ends of the figure which correspond to two 4-d
``branes.'' I have assumed that all the bulk groups
have the same gauge coupling,
consistent with translation invariance in a flat extra
dimension.\footnote{\label{natural}Technically, this is not a natural
constraint. The breakdown of translation invariance at the edges will, in
high enough order, produce infinite $j$-dependent renormalizations of the
bulk couplings, but nobody has worried about that for years.} The links
in the figure are associated with $2\times2$ real $\sigma$ fields satisfying
\begin{equation}
\sigma=s+i\,\vec\tau\cdot\vec p\quad\quad
\sigma_{j,j+1}^*=\tau_2\sigma_{j,j+1}\tau_2
\end{equation}
each with vacuum expectation value $v_0$, again preserving the translation
invariance
\begin{equation}
\left\langle\sigma_{j,j+1}\right\rangle
=v_0
\end{equation}
Note that the VEVs can all be rotated to be proportional to the identity.

Because of the assumption of translation invariance, the gauge
sector of the model is determined by only four parameters, $v_0$, $g_0$,
$g$, and $g_{\np}$, compared to three in the standard model, $v$, $e$ and
$\sin\theta$. 
Though it will not play any 
important role in the tree-level analysis,
it is reasonable to assume a translation invariant set of potentials for
the scalars as well. 

To have fermion delocalization, we need to spread the fermion doublets from
one end of the theory space in figure~\ref{fig-g1} to the other.
The most efficient way to do this, I believe, is to alternate
with the LH doublets on the even nodes and RH doublets on the odd nodes, 
as shown in figure~\ref{fig-g1}. This setup has a number of nice features,
which we will discuss below.
This only works for even $N$, which is why we have assumed that $N$ is even.

The field $\psi^{\np}_R$ is written as a doublet, but because the $N+1$
group is a $U(1)$, the top and bottom components are independent.
\begin{equation}
\psi^{\np}_R=\pmatrix{
U_R\cr
D_R\cr
}
\end{equation}
If we write the fields in a column vector in the theory space
where the LH and RH fields
alternate on the even and odd components,\footnote{For thirty years I
have taught my students not to mix LH and RH fields. Here, for once, it is
appropriate, because the form of the fermion mass matrix ensures that the
couplings have the right form. There is still something odd about the
notation though. Because the fermion states break up into LH states on the
even nodes and RH states on the odd, the eigenvalues come in degenerate
pairs and there is a superselection rule ---
there is no need ever to superpose even-node states with odd-node states.}
\begin{equation}
[\Psi]_{2j}=\psi^{2j}_L\,,\quad
[\Psi]_{2j+1}=\psi^{2j+1}_R\;\;\mbox{for $j=0$ to $N/2$.}
\label{psis}
\end{equation}
and
the Yukawa couplings can be written as
{\renewcommand{\arraystretch}{1.3}\begin{equation}
\begin{array}{c}
\displaystyle
\bar\Psi\,{\cal A}^f\,\Psi\quad\mbox{with}\quad
[{\cal A}^f]_{j,k}=0\;\;\mbox{if $j\neq k\pm1$}\\
\displaystyle
[{\cal A}^f]_{j,j+1}=a_{j,j+1}\sigma_{j,j+1}\,,\;\;
[{\cal A}^f]_{j+1,j}=a_{j,j+1}\sigma_{j,j+1}^\dagger\;\;
\mbox{for $j=0$ to $N$}
\end{array}
\label{generalaf}
\end{equation}}

Because we have imposed translation invariance in the bulk for the gauge
couplings and VEVs, we should also impose translation invariance for the
Yukawa coupling. This, however, is somewhat peculiar given our fermion
representation, because of the alternation of LH and RH fermions in
(\ref{psis}). 
Thus we want our discrete translation invariance in the bulk to have the
form
\begin{equation}
j\to j+1\quad\quad L\leftrightarrow R
\label{fermiontranslations}
\end{equation}
changing parity!. This seems a bit odd, but it has an important 
consequence. Because of (\ref{fermiontranslations}), the fermion
mass matrix in the bulk has the simple form
\begin{equation}
\bar\Psi\,{\cal M}\,\Psi\;\;\mbox{with}\;\;
[{\cal M}]_{j,k}=0\;\;\mbox{if $j\neq k\pm1$}
\quad\quad
[{\cal M}]_{j,j+1}=av_0\,,\;\;
[{\cal M}]_{j+1,j}=av_0
\label{bulkfermions}
\end{equation}
which automatically has zero modes that are spread over the whole
bulk.
I will say more about this below.

We will also assume that the form (\ref{bulkfermions}) is flavor
independent, and in fact that the only flavor dependence is on the $U(1)$
brane at $j=N+1$. 
We will also break the translation invariance on the brane at $j=0$.
We will need this freedom below. But we will assume that the coupling at
the $j=0$ brane is flavor independent.
It might be possible to get
away with very small differences in the ``bulk'' couplings of the various
flavors and the coupling to the $j=0$ brane, 
but this is a very dangerous path, likely to exacerbate the
problem of universality violation and flavor changing neutral currents, and
I want to see how far we can get without walking this plank. 

Making these assumptions, the Yukawa couplings become
{\renewcommand{\arraystretch}{1.3}\begin{equation}
\begin{array}{c}
\displaystyle
\bar\Psi\,{\cal A}^f\,\Psi\quad\mbox{with}\quad
[{\cal A}^f]_{j,k}=0\;\;\mbox{if $j\neq k\pm1$}\\
\displaystyle
[{\cal A}^f]_{j,j+1}=a\,\sigma_{j,j+1}\,,\;\;
[{\cal A}^f]_{j+1,j}=a\,\sigma_{j,j+1}^\dagger\;\;
\mbox{for $j=1$ to $N-1$}\\
\displaystyle
[{\cal A}^f]_{01}=\epsilon_0a\,\sigma_{01}\,,\;\;
[{\cal A}^f]_{10}=\epsilon_0a\,\sigma_{01}^\dagger\\
\displaystyle
[{\cal A}^f]_{N,\np}=\epsilon^{f_U/f_D}_Na\,\sigma_{N,\np}\,,\;\;
[{\cal A}^f]_{\np,N}=\epsilon^{f_U/f_D}_Na\,\sigma_{N,\np}^\dagger
\end{array}
\label{af}
\end{equation}}
with all flavor dependence in the constant $\epsilon^{f_U/f_D}_{N}$.

The mass matrix then has the form
{\renewcommand{\arraystretch}{1.3}\begin{equation}
\begin{array}{c}
\displaystyle
\bar\Psi\,{\cal M}\,\Psi\;\;\mbox{with}\;\;
[{\cal M}]_{j,k}=0\;\;\mbox{if $j\neq k\pm1$}
\quad\quad
[{\cal M}]_{j,j+1}=a\,v_0\,,\;\;
[{\cal M}]_{j+1,j}=a\,v_0
\\
\displaystyle
[{\cal M}]_{01}=\epsilon_0a\,v_0\,,\;\;
[{\cal M}]_{10}=\epsilon_0a\,v_0\,,\;\;
[{\cal M}]_{N,\np}=\epsilon^{f_U/f_D}_Na\,v_0\,,\;\;
[{\cal M}]_{\np,N}=\epsilon^{f_U/f_D}_Na\,v_0
\end{array}
\end{equation}}
As expected, the eigenvectors of this mass matrix come in pairs with
degenerate eigenvalues, eigenvectors $e_L^j$ with only even components and
$e_R^j$ with only odd components.

Note that the only place where there is a
difference between the couplings of the top and bottom
of the RH doublet $\psi^{\np}_R$ is in 
the term proportional to $\epsilon^{f_U/f_D}_{N}$. 
This is always true whatever other
assumptions we make about the form of the Yukawa couplings.

For the light quarks and all the leptons, 
the $\epsilon^{f_U/f_D}_{N}$ constants will be very small,
and to first approximation, we can simply assume 
\begin{equation}
\epsilon^{\rm light\;fermions}_{N}\approx0
\label{anlight}
\end{equation} 
In this limit, the eigenvectors for the LH light modes are very
simple.
For approximately zero eigenvalue, and they all have the form $e_L^0$
where
\begin{equation}
[e_L^0]_{0}=\sqrt{\frac{1}{1+N\epsilon_0^2/2}}\quad\quad
[e_L^0]_{2j}=\alpha_j=(-1)^j\,\epsilon_0\,[e_L^0]_{0}
\quad\quad
[e_L^0]_{2j-1}=0
\label{eL0}
\end{equation}
for $j=1$ to $N/2$.
The reader should now begin to see why we need the extra freedom of
assuming that 
\begin{equation}
\epsilon_0\neq 1
\label{a0neqa}
\end{equation}
The quantity $\epsilon_0$ determines the amount of delocalization
of the LH fermions. In the limit we
are considering, we want to keep the delocalization relatively small so we
are interested in the region
\begin{equation}
|\epsilon_0|\ll 1
\label{a0lla}
\end{equation}
In this limit, the RH light mode is not delocalized. It is stuck on the
``brane'' -- the $N+1$ node
\begin{equation}
[e_R^0]_j=\delta_{j,\np}
\end{equation}
For the $t$ quark, 
however, $\epsilon_N$ is not small, the RH mode is
delocalized and the LH mode is more complicated than (\ref{eL0}).
\cmt{The LH eigenvector $e_L^t$ for the lightest $t$ mode has the form
{\renewcommand{\arraystretch}{1.5}
\begin{equation}
\begin{array}{c}
[e_L^t]_{2j-1}=0\quad\quad [e_L^t]_{2j}=(-1)^j\,[e_L^t]_{0}\times
\\ \displaystyle
\frac{\left(2-\epsilon_0^2-2(1-\epsilon_0^2)\cos c_t\right)
\cos(jc_t+c_t)
-\left(\epsilon_0^2-2(1-\epsilon_0^2)\cos c_t\right)\sin(jc_t+c_t)\tan(c_t/2)
}{\epsilon_0a}
\end{array}
\end{equation}}
for $j=1$ to $N/2$, where
\begin{equation}
c_t=2\arcsin\frac{m_t}{2a\,v_0}
\end{equation}
This satisfies the eigenvalue equation if $(\epsilon_N^t)^2$ is given by
\begin{equation}
\frac{\displaystyle\sin\frac{c_t}{2}\left(8\cos\frac{(N+1)c_t}{2}\sin\frac{c_t}{2}
+4\epsilon_0^2\sin\frac{Nc_t}{2}\right)}
{\displaystyle2\epsilon_0^2\cos\frac{(N-1)c_t}{2}
-4\sin\frac{c_t}{2}\sin\frac{Nc_t}{2}}
\label{ant2}
\end{equation}
Thus we can choose any $\epsilon_0$ and $N$ so long as (\ref{ant2}) gives a
positive value for $(\epsilon_N^t)^2$.}

\cmt{
The mass matrix for the charged gauge bosons is
\begin{equation}
M_c^2=\frac{1}{4}\widetilde G\,\widetilde V\,\widetilde G
\label{mc2}
\end{equation}
where $\widetilde G$ is the $(N+1)\times(N+1)$ 
diagonal matrix of gauge couplings without
$g_{\np}$
\begin{equation}
[\widetilde G]_{jk}=0\;\;\mbox{for $j\neq k$\,,}\quad
[\widetilde G]_{00}=g_0\mbox{\,,}\quad
[\widetilde G]_{jj}=g\;\;\mbox{for $j=1$ to $N$\,.}\quad
\label{tildeg}
\end{equation}
and the matrix $\widetilde V$ is given by
{\renewcommand{\arraystretch}{1.3}
\begin{equation}
\begin{array}{c}
\displaystyle
[\widetilde V]_{jk}=0\;\;\mbox{for $j\neq k,\,k\pm 1$\,,}\quad
[\widetilde V]_{00}=v_0^2\;\;\mbox{\,,}\quad
[\widetilde V]_{jj}=2v_0^2\;\;\mbox{for $j=1$ to $N$\,,}
\\
\displaystyle
[\widetilde V]_{j,j+1}=
[\widetilde V]_{j+1,j}=
-v_0^2\;\;\mbox{for $j=0$ to $N-1$\,.}
\end{array}
\label{tildev}
\end{equation}}
}

The low energy charged-current
weak interactions of the light fermion modes 
are determined by the inverse of $\widetilde
V$ and the eigenvector $e_L^0$:
\begin{equation}
\sqrt{2}\, G_F=\frac{1}{v^2}
=\sum_{j,k=0}^{N+1}[\widetilde V^{-1}]_{jk}|[e_L^0]_{j}|^2|[e_L^0]_{k}|^2
\label{vi}
\end{equation}
\cmt{
The matrix $\widetilde V^{-1}$ satisfies
\begin{equation}
[\widetilde V^{-1}]_{jk}
=\min(N+1-j,N+1-k)/v_0^2
\label{tildevinverse}
\end{equation}
so 
\begin{equation}
\frac{v_0^2}{v^2}
=\frac{N+1+\epsilon_0^2N^2/2+\epsilon_0^4N(N^2+2)/12}
{(1+\epsilon_0^2N^2/2)^2}
\label{v02overv2}
\end{equation}
}
The low energy neutral-current weak interactions are then given in terms of
$\widetilde V^{-1}$ by~\cite{Georgi:1977wk} 
\begin{equation}
\sum_{j,k=0}^{N+1}[\widetilde V^{-1}]_{jk}
\left[T_3|[e_L^0]_j|^2-\frac{e^2}{g_j^2}Q\right]\,
\left[T_3|[e_L^0]_j|^2-\frac{e^2}{g_k^2}Q\right]
\label{nc}
\end{equation}
The normalization of the $T_3^2$ term satisfies custodial
$SU(2)$ symmetry, so the correction to the $\rho$ 
parameter is small. The analog of
$\sin^2\theta$ as determined by the low energy weak interactions is
determined by the coefficient of $T_3\, Q$ in (\ref{nc}) to satisfy
\begin{equation}
\sqrt{2}G_F\sin^2\theta
=
\sum_{j,k=0}^{N}[\widetilde V^{-1}]_{jk}
|[e_L^0]_j|^2\;\frac{e^2}{g_k^2}
\label{sin2theta}
\end{equation}
\cmt{
which using (\ref{eL0}) and (\ref{tildevinverse}), we can write as
\begin{equation}
\frac{(N+1)e^2/g_0^2+\epsilon_0^2 N^2e^2/4g_0^2
+N(N+1)e^2/2g^2
+\epsilon_0^2N(4 N^2+3 N+2)e^2/24g^2}
{1+ \epsilon_0^2N/2}
\label{sin2thetaexplicit}
\end{equation}
}

\cmt{
The mass matrix for the neutral gauge bosons is
\begin{equation}
M_n^2=\frac{1}{4} G\, V\, G
\label{mn2}
\end{equation}
where $ G$ is the $(N+2)\times(N+2)$ 
diagonal matrix of gauge couplings
\begin{equation}
[G]_{jk}=0\;\;\mbox{for $j\neq k$\,,}\quad
[G]_{00}=g_0\mbox{\,,}\quad
[G]_{jj}=g\;\;\mbox{for $j=1$ to $N$\,,}\quad
[G]_{\np,\np}=g_{\np}\mbox{\,.}
\label{g}
\end{equation}
and the matrix $ V$ is
{\renewcommand{\arraystretch}{1.3}
\begin{equation}
\begin{array}{c}
\displaystyle
[V]_{jk}=0\;\;\mbox{for $j\neq k,\,k\pm 1$\,,}\quad
[V]_{jj}=2v_0^2\;\;\mbox{for $j=1$ to $N$\,,}
\\
\displaystyle
[V]_{00}=
[V]_{\np,\np}=
v_0^2\;\;\mbox{\,,}\quad
[V]_{j,j+1}=
[V]_{j+1,j}=
-v_0^2\;\;\mbox{for $j=0$ to $N$\,.}
\end{array}
\label{v}
\end{equation}}
The neutral mass squared matrix given by (\ref{mn2}) has, of
course, a zero eigenvalue associated with the photon. 
The photon eigenstate $\kappa_{\np}$ is given by
\begin{equation}
[\kappa^{\np}]_0=\frac{e}{g_0}\;,\quad\quad
[\kappa^{\np}]_j=\frac{e}{g}\quad\mbox{for $j=1$ to $N$,}\quad\quad
[\kappa^{\np}]_{\np}=\frac{e}{g_{\np}}
\end{equation}
}

Having specified the model in general, let us now look at what happens for
a particular set of parameters chosen to produce something like the standard
model at low energies. To produce these values, I fixed $e$, $v$, $M_W$,
$M_Z$ and $m_t$, 
and then scanned over various values of $\epsilon_0$, $a$ and $N$. To do
this efficiently, it is convenient to have analytic expressions for $g_0$,
$g_{\np}$ and $\epsilon_{\np}^t$ in terms of the fixed values. This is
possible because of the translation invariance in bulk, and this and other
details of the search procedure will be discussed fully in a forthcoming
paper~\cite{longer}. 
But it is straightforward, given a set of parameters, $N$, $g_0$,
$g$, $g_{\np}$, $v_0$, $a$, $\epsilon_0$ and $\epsilon_{\np}^t$ to
diagonalize the mass matrices 
numerically and see that it all works.\footnote{Of
course, it was less straightforward in the early 70s, when it was a major
production to do numerical work on anything more powerful than a
slide-rule.} Here is one set of parameters:
{\renewcommand{\arraystretch}{1.2}
\begin{equation}
\begin{array}{c}
N=88\,,\;\;
g=7.8\,,\;\;
g_0=0.899\,,\;\;
g_{\np}=0.363\,,\;\;
\\
v_0=2.005\,\mbox{TeV}\,,\;\;
a=7\,,\;\;
\epsilon_0=0.1\,,\;\;
\epsilon_{\np}^t=0.309\,.
\end{array}
\label{parameters}
\end{equation}}%
These give
{\renewcommand{\arraystretch}{1.2}
\begin{equation}
\begin{array}{c}
\alpha = 1/129\,,\;\;
v = 250\,\mbox{GeV}\,,\;\;
\sin^2\theta=0.228
\\
M_W = 80.425\,\mbox{GeV}\,,\;\;
M_Z = 91.1876\,\mbox{GeV}\,,\;\;
m_t = 175\,\mbox{GeV}\,,\;\;
\\ \displaystyle
\rule[-2.5ex]{0ex}{7ex}\frac{W{LL}}{{\rm s.m.}}=0.987\,,\;\;
\frac{W{t_Lb_L}}{{\rm s.m.}}=1.052\,,\;\;
\frac{Z{LL}}{{\rm s.m.}}=0.985\,,\;\;
\frac{Z{RR}}{{\rm s.m.}}=0.972\,,\;\;
\\ \displaystyle
\rule[-2.5ex]{0ex}{7ex}\frac{WWZ}{{\rm s.m.}}=1.097\,,\;\;
\frac{Zt_Lt_L}{{\rm s.m.}}=1.137\,,\;\;
\frac{Zt_Rt_R}{{\rm s.m.}}=0.096\,.\;\;
\end{array}
\label{results}
\end{equation}}%
In (\ref{results}), $\sin^2\theta$ is the tree-level value in the
low-energy neutral-current weak interactions, from (\ref{sin2theta}). The
couplings of the $W$ and $Z$ are tabulated using the numerical eigenvectors
and compared to the tree-level values in the standard model.\footnote{$L$
($R$) stands for any LH (RH) light field.}

In table~\ref{table-1}, I tabulate the masses and some of the couplings of
the first nine recurrences of the $W$ and $Z$. Table~\ref{table-2}, 
contains similar information for 
the first nine recurrences of the light fermions and the $t$.
{\renewcommand{\arraystretch}{1.1}\figsize\begin{table}[htb]
$$
\begin{array}{|c|c|c|c||c|c|c|c|c|c|}
\hline
\displaystyle {M_W\atop\mbox{(GeV)}}& 
\displaystyle\rule[-2.5ex]{0ex}{6.5ex}\frac{W{LL}}{{\rm s.m.}}&
\displaystyle\frac{W{t_Lb_L}}{{\rm s.m.}}& 
\displaystyle\frac{W_{k}WZ}{{\rm s.m.}}& 
\displaystyle {M_Z\atop\mbox{(GeV)}}&
\displaystyle\frac{Z{LL}}{{\rm s.m.}}& 
\displaystyle\frac{Z{RR}}{{\rm s.m.}}& 
\displaystyle\frac{Z{t_Lt_L}}{{\rm s.m.}}& 
\displaystyle\frac{Z{t_Rt_R}}{{\rm s.m.}}& 
\displaystyle\frac{Z_{k}WW}{{\rm s.m.}}\\ 
\hline
\hline
80.42& 0.99& 1.036& 1.097& 91.19& 0.988& 0.973& 1.137& 0.961& 1.097\\
\hline
305.03& 0.088& 0.239& 0.224& 309.46& 0.099& 0.357& 0.382& 2.871& 0.182\\
\hline
567.93& 0.224& 0.173& 0.012& 570.47& 0.254& 0.2& 0.197& 0.334& 0.011\\
\hline
838.58& 0.038& 0.093& 0.008& 840.33& 0.043& 0.137& 0.132& 1.05& 0.007\\
\hline
1111.41& 0.116& 0.09& 0.001& 1112.74& 0.133& 0.104& 0.103& 0.168& 0.001\\
\hline
1384.94& 0.023& 0.057& 0.002& 1386.& 0.026& 0.083& 0.079& 0.635& 0.002\\
\hline
1658.57& 0.078& 0.06& \approx0& 1659.45& 0.089& 0.069& 0.069& 0.109& \approx0\\
\hline
1931.99& 0.017& 0.041& 0.001& 1932.75& 0.019& 0.059& 0.057& 0.453& 0.001\\
\hline
2205.& 0.059& 0.045& \approx0& 2205.67& 0.068& 0.052& 0.052& 0.078& \approx0\\
\hline
2477.46& 0.013& 0.032& \approx0& 2478.05& 0.015& 0.046& 0.044& 0.35& \approx0\\
\hline
\end{array}$$
\caption{\figsize\sf\label{table-1}The $W$ and
$Z$ and their first nine recurrences, along with their couplings to quarks
compared to the standard model couplings.}\end{table}}
{\renewcommand{\arraystretch}{1.1}\figsize\begin{table}[htb]
$$
\begin{array}{|c|c|c|c|c||c|c|c|c|}
\hline
m\mbox{~(GeV)}& 
\displaystyle\rule[-2.5ex]{0ex}{6.5ex}\frac{W{u_kd_1}}{{\rm s.m.}}&
\displaystyle\frac{W{t_1b_k}}{{\rm s.m.}}&
\displaystyle\frac{Z{L_kL_1}}{{\rm s.m.}}&
\displaystyle\frac{Z{R_kR_1}}{{\rm s.m.}}& m_t\mbox{~(GeV)}&
\displaystyle\frac{W{t_kb_1}}{{\rm s.m.}}&
\displaystyle\frac{Z{t_kt_1}L}{{\rm s.m.}}&
\displaystyle\frac{Z{t_kt_1R}}{{\rm s.m.}}\\ 
\hline
\hline
571.72& 0.696& 0.516& 0.769& \approx0& 867.42& 0.738& 0.59& 0.256\\
\hline
1514.91& 0.551& 0.409& 0.617& \approx0& 1726.51& 0.662& 0.518& 1.183\\
\hline
2491.81& 0.512& 0.387& 0.576& \approx0& 2643.94& 0.613& 0.49& 1.582\\
\hline
3472.06& 0.495& 0.378& 0.559& \approx0& 3587.82& 0.58& 0.473& 1.74\\
\hline
4450.21& 0.485& 0.374& 0.549& \approx0& 4542.45& 0.558& 0.462& 1.785\\
\hline
5423.85& 0.479& 0.371& 0.543& \approx0& 5499.85& 0.541& 0.454& 1.772\\
\hline
6391.3& 0.475& 0.369& 0.538& \approx0& 6455.46& 0.529& 0.448& 1.726\\
\hline
7351.12& 0.472& 0.368& 0.535& \approx0& 7406.28& 0.519& 0.443& 1.658\\
\hline
8302.01& 0.469& 0.367& 0.533& \approx0& 8350.07& 0.512& 0.439& 1.577\\
\hline
\end{array}$$
\caption{\figsize\sf\label{table-2}The first nine recurrences for light
quarks and the $t$, along with some of their couplings to light quarks
and $W$ and $Z$, compared to the standard model couplings.}\end{table}}
There is interesting phenomenology here beyond what already appears in the
literature~\cite{Barbieri:2003pr,Birkedal:2004au,Chanowitz:2004gk}. For
example, in this class of models, splittings between fermion recurrences are 
much larger than between gauge boson recurrences 
because $a>g/2$ and because
the chiral delocalization effectively reduces the size of
the extra dimension for the fermions in half.\footnote{I am grateful the
Sekhar Chivukula for pointing out the effective size difference.} 

Now for some comments
\begin{enumerate}
\item Because we have chosen parameters to give the right masses and
couplings for $W$ and $Z$, we automatically ensure that the $S$ parameter
is small and that other low energy tests of the standard model are
satisfied. The couplings of the $t$ and the $WWZ$ couplings show more
deviation from the tree-level standard model, as one would expect because
the effective theory at large energies is very different. 
Clearly, the
values in (\ref{parameters}) are
finely tuned to produce something that looks like the standard
model. On the other hand, this model does not have a lot of parameters to
tune. It is 
surprising, at least to me, that one can do this at all. In particular, we
are able to get a large enough $t$ mass without making the $Zbb$ couplings
significantly different from those of the other light quarks. This has been
a worry in previous works~\cite{Csaki:2003sh,Cacciapaglia:2005pa,
Schwinn:2005qa}.
\item Some of the couplings are quite large, and one certainly has to worry
that loop-corrections will modify things.  
\item To go beyond the $2$~TeV level for the VEVs in this class of
models requires even larger Yukawa and gauge couplings. But even if one
does not worry about the size of the couplings, at some point, this program
runs out of steam, because there is no 
way to get the appropriate gauge boson and $t$ masses with real
couplings. 
\item Since there are lots of parameters, discrepancies at the percent
level are interesting only if there are bounds that force them to go in a
particular direction. That is probably the case for the couplings,
$W{LL}$, $Z{LL}$ and $Z{RR}$, which are systematically smaller than the
tree-level standard model, presumably because some of the low-energy weak
interactions arise from coupling of the light fermions to heavy gauge
bosons. This model does not have ``ideal
delocalization''~\cite{Chivukula:2005xm} in which these couplings
vanish. My belief is that it is very difficult to implement this in any
natural way so I find it heartening that at least these discrepancies can
be made quite small by a suitable choice of parameters.
\item Searching even this small parameter space is greatly facilitated by
having analytic expressions for many of the parameters, which is possible
because of the translation invariance of the bulk. Very likely there are
slightly warped models~\cite{Nomura:2003du,Davoudiasl:2003me,
Cacciapaglia:2004jz,Davoudiasl:2004pw,Hewett:2004dv} 
nearby with even better correspondence
to the standard model. And of course, there may be strongly warped
solutions in very different regions of parameter space --- I just 
do not know how to look for them efficiently.
\item The most interesting thing, I think, about this little exercise, is
the unusual assignment of fermions to nodes in (\ref{psis}). While this
makes perfect sense in the 4-d picture, it is not obvious what it means for
the interpretation of the deconstructed extra dimension. It is hard to see
how color, for example, could be spread over such a construction. 
\end{enumerate}

\section*{Acknowledgements}
I am grateful for useful discussions with Sekhar Chivukula, Csaba Csaki and 
Daniele Dominici. This research is supported in part by
the National Science Foundation under grant PHY-0244821.

\bibliography{higgsless2}

\providecommand{\href}[2]{#2}\begingroup\raggedright\begin{thebibliography}{10}

\bibitem{Csaki:2003dt}
C.~Csaki, C.~Grojean, H.~Murayama, L.~Pilo, and J.~Terning, ``Gauge theories on
  an interval: Unitarity without a higgs,'' {\em Phys. Rev.} {\bf D69} (2004)
  055006,
\href{http://www.arXiv.org/abs/hep-ph/0305237}{{\tt hep-ph/0305237}}.

\bibitem{Csaki:2003zu}
C.~Csaki, C.~Grojean, L.~Pilo, and J.~Terning, ``Towards a realistic model of
  higgsless electroweak symmetry breaking,'' {\em Phys. Rev. Lett.} {\bf 92}
  (2004) 101802,
\href{http://www.arXiv.org/abs/hep-ph/0308038}{{\tt hep-ph/0308038}}.

\bibitem{Arkani-Hamed:2001ca}
N.~Arkani-Hamed, A.~G. Cohen, and H.~Georgi, ``(de)constructing dimensions,''
  {\em Phys. Rev. Lett.} {\bf 86} (2001) 4757--4761,
\href{http://www.arXiv.org/abs/hep-th/0104005}{{\tt hep-th/0104005}}.

\bibitem{Hill:2000mu}
C.~T. Hill, S.~Pokorski, and J.~Wang, ``Gauge invariant effective lagrangian
  for kaluza-klein modes,'' {\em Phys. Rev.} {\bf D64} (2001) 105005,
\href{http://www.arXiv.org/abs/hep-th/0104035}{{\tt hep-th/0104035}}.

\bibitem{Foadi:2003xa}
R.~Foadi, S.~Gopalakrishna, and C.~Schmidt, ``Higgsless electroweak symmetry
  breaking from theory space,'' {\em JHEP} {\bf 03} (2004) 042,
\href{http://www.arXiv.org/abs/hep-ph/0312324}{{\tt hep-ph/0312324}}.

\bibitem{Hirn:2004ze}
J.~Hirn and J.~Stern, ``The role of spurions in higgs-less electroweak
  effective theories,'' {\em Eur. Phys. J.} {\bf C34} (2004) 447--475,
\href{http://www.arXiv.org/abs/hep-ph/0401032}{{\tt hep-ph/0401032}}.

\bibitem{Casalbuoni:2004id}
R.~Casalbuoni, S.~De~Curtis, and D.~Dominici, ``Moose models with vanishing s
  parameter,'' {\em Phys. Rev.} {\bf D70} (2004) 055010,
\href{http://www.arXiv.org/abs/hep-ph/0405188}{{\tt hep-ph/0405188}}.

\bibitem{Chivukula:2004pk}
R.~S. Chivukula, E.~H. Simmons, H.-J. He, M.~Kurachi, and M.~Tanabashi, ``The
  structure of corrections to electroweak interactions in higgsless models,''
  {\em Phys. Rev.} {\bf D70} (2004) 075008,
\href{http://www.arXiv.org/abs/hep-ph/0406077}{{\tt hep-ph/0406077}}.

\bibitem{Evans:2004rc}
N.~Evans and P.~Membry, ``Higgless w unitarity from decoupling
  deconstruction,''
\href{http://www.arXiv.org/abs/hep-ph/0406285}{{\tt hep-ph/0406285}}.

\bibitem{Georgi:2004iy}
H.~Georgi, ``Fun with higgsless theories,'' {\em Phys. Rev.} {\bf D71} (2005)
  015016,
\href{http://www.arXiv.org/abs/hep-ph/0408067}{{\tt hep-ph/0408067}}.

\bibitem{Chivukula:2004af}
R.~S. Chivukula, E.~H. Simmons, H.-J. He, M.~Kurachi, and M.~Tanabashi,
  ``Universal non-oblique corrections in higgsless models and beyond,'' {\em
  Phys. Lett.} {\bf B603} (2004) 210--218,
\href{http://www.arXiv.org/abs/hep-ph/0408262}{{\tt hep-ph/0408262}}.

\bibitem{Kurachi:2004rj}
M.~Kurachi, R.~S. Chivukula, E.~H. Simmons, H.-J. He, and M.~Tanabashi,
  ``Oblique corrections in deconstructed higgsless models,''
\href{http://www.arXiv.org/abs/hep-ph/0409134}{{\tt hep-ph/0409134}}.

\bibitem{Chivukula:2004mu}
R.~S. Chivukula, E.~H. Simmons, H.-J. He, M.~Kurachi, and M.~Tanabashi,
  ``Electroweak corrections and unitarity in linear moose models,'' {\em Phys.
  Rev.} {\bf D71} (2005) 035007,
\href{http://www.arXiv.org/abs/hep-ph/0410154}{{\tt hep-ph/0410154}}.

\bibitem{Chivukula:2005bn}
R.~S. Chivukula, E.~H. Simmons, H.-J. He, M.~Kurachi, and M.~Tanabashi,
  ``Deconstructed higgsless models with one-site delocalization,'' {\em Phys.
  Rev.} {\bf D71} (2005) 115001,
\href{http://www.arXiv.org/abs/hep-ph/0502162}{{\tt hep-ph/0502162}}.

\bibitem{Cacciapaglia:2004rb}
G.~Cacciapaglia, C.~Csaki, C.~Grojean, and J.~Terning, ``Curing the ills of
  higgsless models: The s parameter and unitarity,'' {\em Phys. Rev.} {\bf D71}
  (2005) 035015,
\href{http://www.arXiv.org/abs/hep-ph/0409126}{{\tt hep-ph/0409126}}.

\bibitem{Foadi:2004ps}
R.~Foadi, S.~Gopalakrishna, and C.~Schmidt, ``Effects of fermion localization
  in higgsless theories and electroweak constraints,'' {\em Phys. Lett.} {\bf
  B606} (2005) 157--163,
\href{http://www.arXiv.org/abs/hep-ph/0409266}{{\tt hep-ph/0409266}}.

\bibitem{Casalbuoni:2005rs}
R.~Casalbuoni, S.~De~Curtis, D.~Dolce, and D.~Dominici, ``Playing with fermion
  couplings in higgsless models,'' {\em Phys. Rev.} {\bf D71} (2005) 075015,
\href{http://www.arXiv.org/abs/hep-ph/0502209}{{\tt hep-ph/0502209}}.

\bibitem{Chivukula:2005xm}
R.~S. Chivukula, E.~H. Simmons, H.-J. He, M.~Kurachi, and M.~Tanabashi, ``Ideal
  fermion delocalization in higgsless models,'' {\em Phys. Rev.} {\bf D72}
  (2005) 015008,
\href{http://www.arXiv.org/abs/hep-ph/0504114}{{\tt hep-ph/0504114}}.

\bibitem{Georgi:1977wk}
H.~Georgi and S.~Weinberg, ``Neutral currents in expanded gauge theories,''
  {\em Phys. Rev.} {\bf D17} (1978)
275.

\bibitem{longer}
H.~Georgi, ``in preparation,''.

\bibitem{Barbieri:2003pr}
R.~Barbieri, A.~Pomarol, and R.~Rattazzi, ``Weakly coupled higgsless theories
  and precision electroweak tests,'' {\em Phys. Lett.} {\bf B591} (2004)
  141--149,
\href{http://www.arXiv.org/abs/hep-ph/0310285}{{\tt hep-ph/0310285}}.

\bibitem{Birkedal:2004au}
A.~Birkedal, K.~Matchev, and M.~Perelstein, ``Collider phenomenology of the
  higgsless models,'' {\em Phys. Rev. Lett.} {\bf 94} (2005) 191803,
\href{http://www.arXiv.org/abs/hep-ph/0412278}{{\tt hep-ph/0412278}}.

\bibitem{Chanowitz:2004gk}
M.~S. Chanowitz, ``The no-higgs signal: Strong w w scattering at the lhc,''
\href{http://www.arXiv.org/abs/hep-ph/0412203}{{\tt hep-ph/0412203}}.

\bibitem{Csaki:2003sh}
C.~Csaki, C.~Grojean, J.~Hubisz, Y.~Shirman, and J.~Terning, ``Fermions on an
  interval: Quark and lepton masses without a higgs,'' {\em Phys. Rev.} {\bf
  D70} (2004) 015012,
\href{http://www.arXiv.org/abs/hep-ph/0310355}{{\tt hep-ph/0310355}}.

\bibitem{Cacciapaglia:2005pa}
G.~Cacciapaglia, C.~Csaki, C.~Grojean, M.~Reece, and J.~Terning, ``Top and
  bottom: A brane of their own,''
\href{http://www.arXiv.org/abs/hep-ph/0505001}{{\tt hep-ph/0505001}}.

\bibitem{Schwinn:2005qa}
C.~Schwinn, ``Unitarity constraints on top quark signatures of higgsless
  models,'' {\em Phys. Rev.} {\bf D71} (2005) 113005,
\href{http://www.arXiv.org/abs/hep-ph/0504240}{{\tt hep-ph/0504240}}.

\bibitem{Nomura:2003du}
Y.~Nomura, ``Higgsless theory of electroweak symmetry breaking from warped
  space,'' {\em JHEP} {\bf 11} (2003) 050,
\href{http://www.arXiv.org/abs/hep-ph/0309189}{{\tt hep-ph/0309189}}.

\bibitem{Davoudiasl:2003me}
H.~Davoudiasl, J.~L. Hewett, B.~Lillie, and T.~G. Rizzo, ``Higgsless
  electroweak symmetry breaking in warped backgrounds: Constraints and
  signatures,'' {\em Phys. Rev.} {\bf D70} (2004) 015006,
\href{http://www.arXiv.org/abs/hep-ph/0312193}{{\tt hep-ph/0312193}}.

\bibitem{Cacciapaglia:2004jz}
G.~Cacciapaglia, C.~Csaki, C.~Grojean, and J.~Terning, ``Oblique corrections
  from higgsless models in warped space,'' {\em Phys. Rev.} {\bf D70} (2004)
  075014,
\href{http://www.arXiv.org/abs/hep-ph/0401160}{{\tt hep-ph/0401160}}.

\bibitem{Davoudiasl:2004pw}
H.~Davoudiasl, J.~L. Hewett, B.~Lillie, and T.~G. Rizzo, ``Warped higgsless
  models with ir-brane kinetic terms,'' {\em JHEP} {\bf 05} (2004) 015,
\href{http://www.arXiv.org/abs/hep-ph/0403300}{{\tt hep-ph/0403300}}.

\bibitem{Hewett:2004dv}
J.~L. Hewett, B.~Lillie, and T.~G. Rizzo, ``Monte carlo exploration of warped
  higgsless models,'' {\em JHEP} {\bf 10} (2004) 014,
\href{http://www.arXiv.org/abs/hep-ph/0407059}{{\tt hep-ph/0407059}}.

\end{thebibliography}\endgroup

\end{document}